\documentclass[aps,prb,twocolumn,superscriptaddress]{revtex4-1}
\usepackage[a4paper,top=0.85in,bottom=0.9in,left=0.75in,right=0.75in]{geometry}
\usepackage{amsmath}
\usepackage{amsthm}
\usepackage{amssymb}
\usepackage{latexsym}
\usepackage{bm}
\usepackage{dcolumn}
\usepackage{braket}
\usepackage{graphicx}
\usepackage{epstopdf}
\usepackage{color}
\DeclareGraphicsExtensions{.eps, .pdf, .png}
\usepackage[pdftex]{hyperref}
\definecolor{ltgray}{rgb}{0.95,0.95,0.95}
\definecolor{cream}{rgb}{1,1,0.7}
\definecolor{orange}{rgb}{1,0.644,0}

\newcommand{\He}{$^3$He}
\newcommand{\Hea}{$^3$He-A}
\newcommand{\Heb}{$^3$He-B}
\newcommand{\Hefour}{$^4$He}
\newcommand{\sio}{SiO$_2$}
\def\vp{{\bf p}} 
\newcommand{\detail}[1]
           {\begin{center}\fcolorbox{orange}{ltgray}{
	   \begin{minipage}{0.915\hsize}\small\baselineskip=1pt #1 \end{minipage}
	    }
	    \end{center}}
\def\nicefrac#1#2{\genfrac{}{}{}{1}{#1}{#2}}
\begin{document}
\title{New Phases of Superfluid $^3$He Confined in Aerogels}
\date{\today}
\author{W. P. Halperin}
\affiliation{Department of Physics, Northwestern University, Evanston, IL 60208}
\author{J. M. Parpia}
\affiliation{Department of Physics, Cornell University, Ithaca, NY 14853}
\author{J. A. Sauls}
\affiliation{Center for Applied Physics \& Superconducting Technologies, 
             Northwestern University, Evanston, IL 60208}
\affiliation{Department of Physics, Northwestern University, Evanston, IL 60208}
\begin{abstract}
Liquid $^3$He confined in low-density, highly porous random solids such as silica aerogel provides tuneable systems to study the effects of disorder and confinement on the properties of a quantum liquid. New superfluid phases result from the interplay between disorder, confinement and complex symmetry-breaking.
An extended bibliography is appended.
An edited and abbreviated version of this article appeared in {\sl Physics Today 71, 11, 30 (2018)}: DOI: 
\href{http://dx.doi.org\/10.1063/PT.3.4067}{10.1063/PT.3.4067}. 
\end{abstract}
\maketitle

This article highlights the remarkable new superfluid phases of liquid Helium-Three (\He\ - the light isotope of Helium) that result from confinement of \He\ within highly porous random solids. The orginial discoveries were with \He\ infused into a form of glass (\sio) called silica aerogel. The latter is a remarkable material itself, a gossamer solid with fractal correlations and a density that is $\approx 1/100^{\mbox{th}}$ that of everyday glass.\cite{por99} As for the Helium liquids they are among nature's finer gifts to physics, in part because they exist in the liquid state down to the absolute zero of temperature. This has provided us with a liquid whose properties in aggregate are governed by the laws of quantum mechanics, i.e the Helium liquids are quantum liquids, a field of inquiry pioneered by Fritz London.~\cite{london54}

In 1971 the transitions to the superfluid phases of liquid \He\ were detected by Pomeranchuk cooling along the melting curve by the Cornell group.~\cite{osh72} That landmark discovery was the realization of the Bardeen, Cooper and Schrieffer (BCS) pairing theory of superconductivity in a quantum liquid. Soon after their discovery, the superfluids were identified based in large part on the theory of nuclear magnetic resonance (NMR) developed by Anthony Leggett for BCS condensates of Cooper pairs, each with orbital angular momentum $L=\hbar$ (orbital p-wave) and total spin $S=\hbar$ (spin triplet).~\cite{leg73a} 

The discovery of superfluidity of liquid \He\ infused into silica aerogel opened new directions in the study of disorder on BCS condensates with complex broken symmetry phases. Prior to 1995 it was unclear whether or not the p-wave orbital correlations would survive the random potential of a disordered solid like that of silica aerogel.
With high-porosity silica aerogel obtained from the laboratory of Moses Chan at Penn State the first observations of a superfluid state of \He\ infused into silica aerogel were reported by Jeevak Parpia's lab at Cornell using a mechanical oscillator to directly measure the superfluid mass fraction (Fig. \ref{PhaseDiagram}a), and Bill Halperin's lab at Northwestern using NMR to detect the superfluid contribution to the resonance frequency shift (Fig. \ref{PhaseDiagram}b).~\cite{por95,spr95}
The transitions are sharply defined, as indicated by the heat capacity signature in Fig. \ref{PhaseDiagram}c, altogether demonstrating the existence of a phase-coherent superfluid. Early experiments hinted at fundamental changes in the nature of the superfluid phase compared to that of pure \He, notably (i) the absence of a second phase at high pressures, (ii) the appearance of a disorder-induced quantum phase transition at a critical pressure of $p_c\approx 5$ bar (Fig. \ref{PhaseDiagram}d), below which superfluidity is suppressed completely,~\cite{mat97} and (iii) the puzzling observation of an equal-spin pairing (ESP) phase onsetting at $T_c$ for pressures well below the bulk poly-critical point (PCP).\cite{spr95}
Much more research has been carried out in many laboratories since the initial discoveries. In what follows we highlight some of the research on \He\ infused into high-porosity disordered solids, particularly the discovery and control of new superfluid phases.

\vspace*{-5mm}
\subsection*{Background: Pure \He}
\vspace*{-3mm}

Over a wide range of temperatures below the Fermi temperature, $2\,\mbox{mK} \lesssim T \lesssim 1\,\mbox{K}$, liquid $^3$He is well described by the Landau Fermi-liquid theory, formulated in terms low-energy fermionic quasiparticles. The thermodynamic and transport properties of the normal Fermi liquid are determined by these excitations, their interactions, and the resulting composite bosonic excitations, particularly ``zero-sound'' phonons and long-lived spin excitations, or ``paramagnons''. 
The latter are responsible for the binding of pairs of spin $s=\nicefrac{1}{2}$ quasiparticles into Cooper pairs, and the instability of the Fermi liquid to the formation of a condensate of composite bosons, Cooper pairs, with spin $S=1$.\cite{vollhardt90} The instability occurs at low temperatures, $T_c = 1.0-2.5\,\mbox{mK}$ depending on pressure, as a second-order phase transition to one of two superfluid phases, the B phase for pressures $p<p_c\approx 21\,\mbox{bar}$, or the A phase for $p>p_c$. A first-order transition from the A- to the B-phase is observed along the line $T_{\mbox{\tiny AB}}(p)$ shown as the dashed red line in Fig. \ref{PhaseDiagram}d. The two transition lines converge at a critical point $T_{\mbox{\tiny AB}}(p_c)=T_c(p_c)$, indicated by ``PCP''. Both A- and B phases are definitively identified as BCS condensates of spin-triplet ($S=1$), orbital p-wave ($L=1$) Cooper pairs.
The original observations led to the Nobel prize in 1996 to Douglas Osheroff, Robert Richardson and David Lee, and in 2003 to Anthony Leggett for his theory of NMR for spin-triplet BCS condensates that was developed in parallel with the experimental investigations, and was instrumental identifying the symmetries of the phases.\cite{leg72} 

\begin{widetext}
\begin{minipage}{\textwidth}
\includegraphics[width=0.45\textwidth]{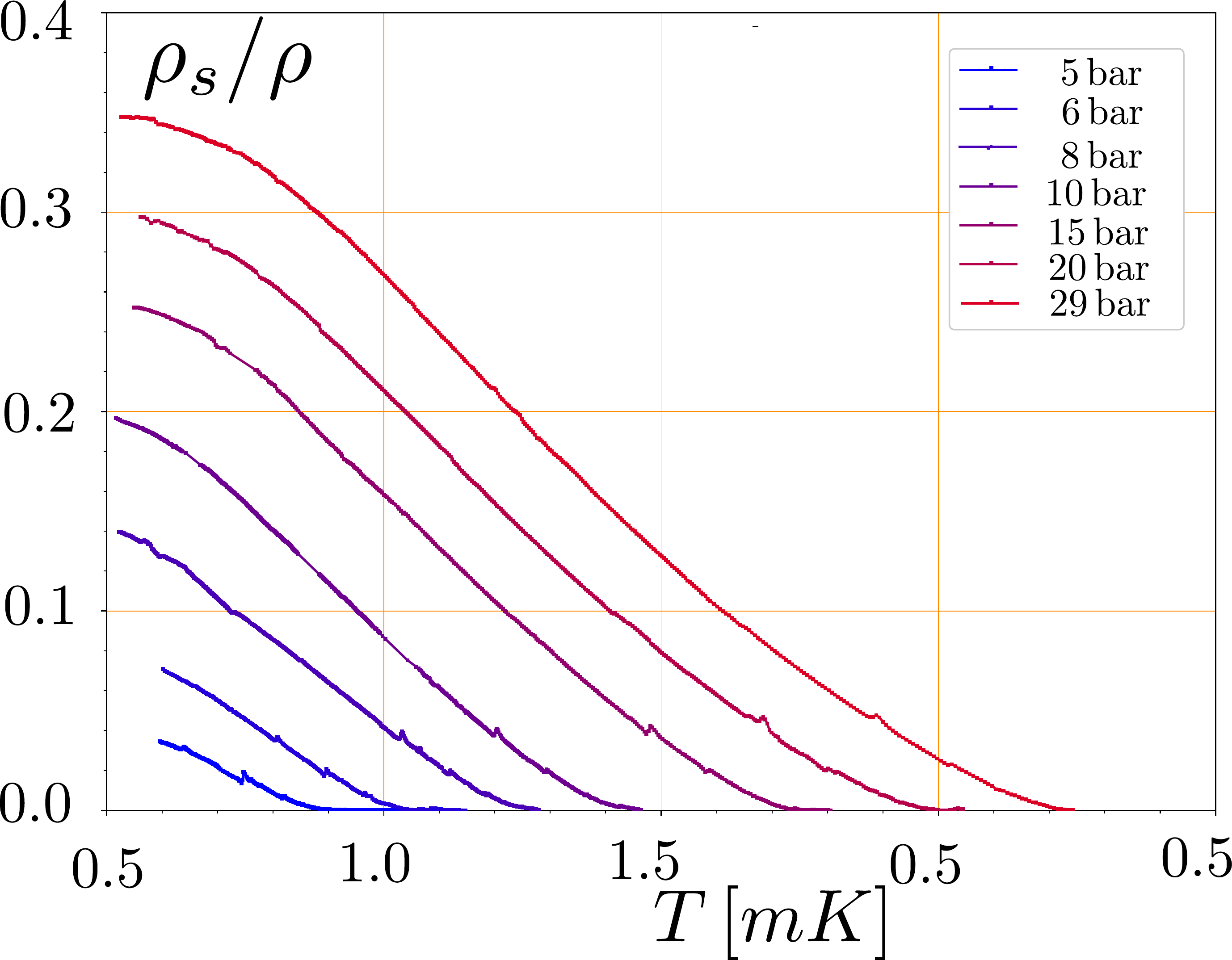}
\includegraphics[width=0.5\textwidth]{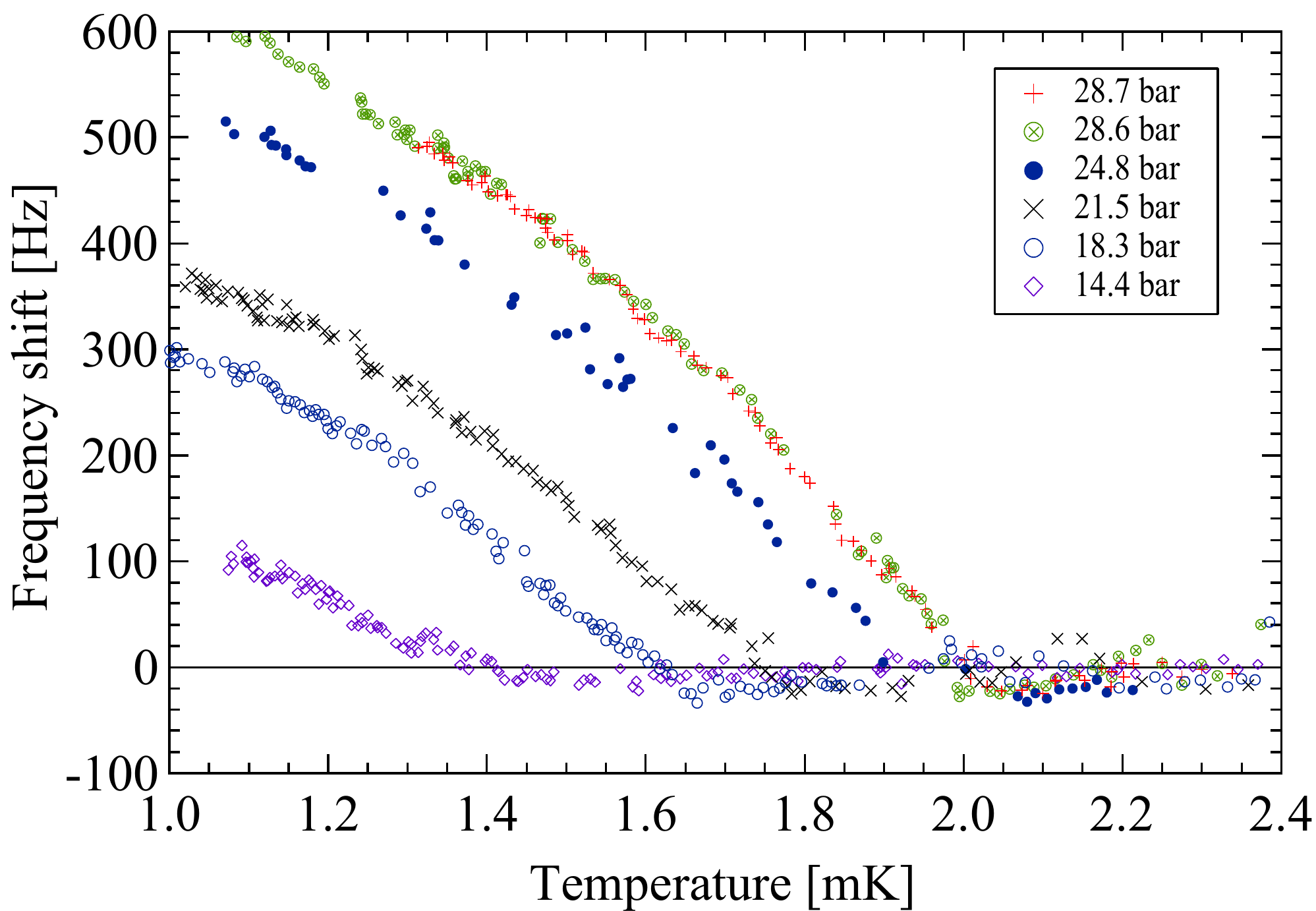}
\end{minipage}
\begin{minipage}{\textwidth}
\hspace*{5mm}\includegraphics[width=0.45\textwidth]{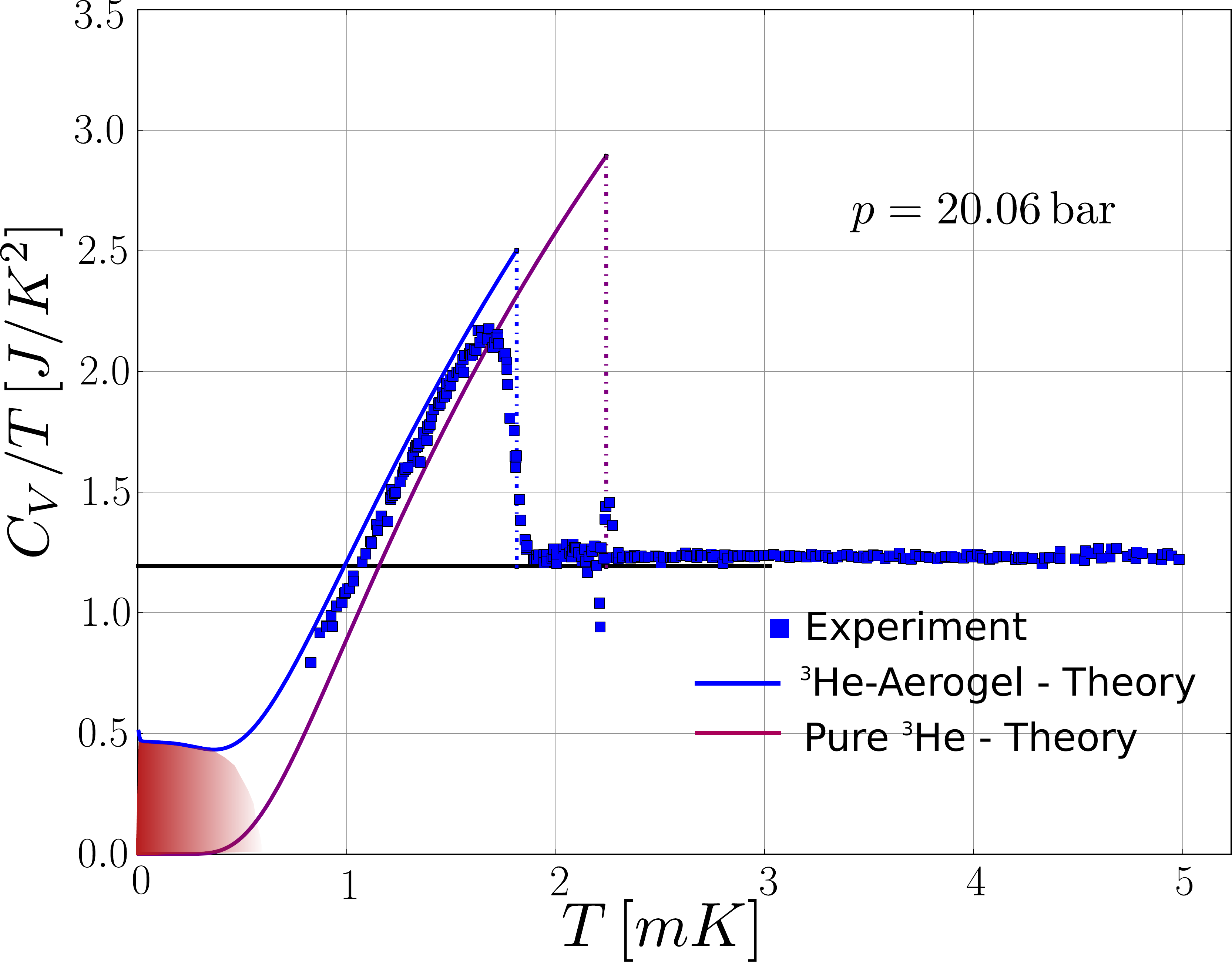}
\hspace*{5mm}\includegraphics[width=0.45\textwidth]{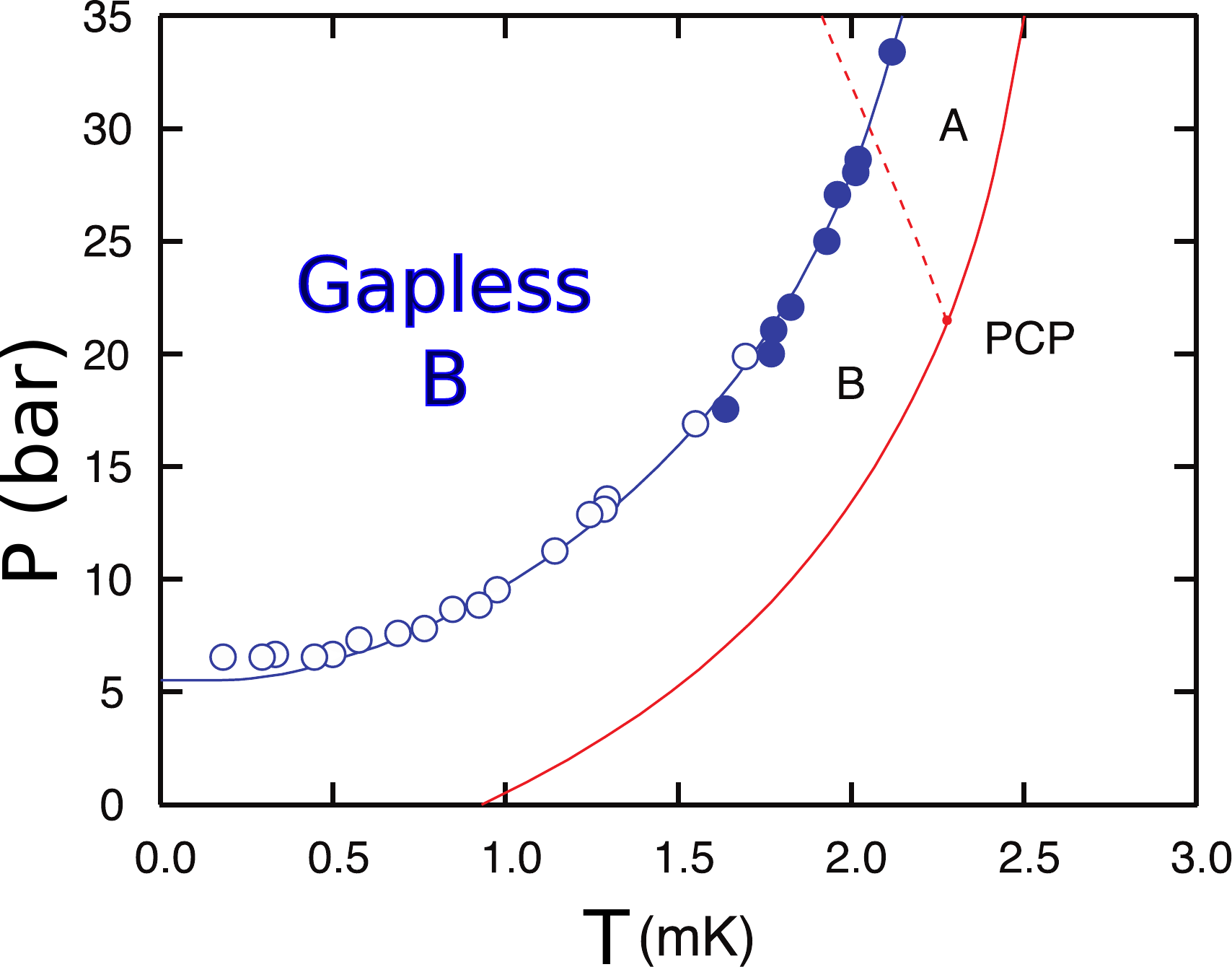}
\end{minipage}
\vspace*{-5mm}
\begin{figure}[h]
\caption{\label{PhaseDiagram} 
Measurements and data showing signatures of the superfluid phase of \He\ in 98\% porous silica aerogel, and comparisons with bulk superfluid \He\ as a function of temperature. (a) superfluid fraction $\rho_s/\rho$: the inset shows the suppression relative to pure \He\ at $p=29$ bar, (b) NMR frequency shift, $\Delta\omega$, (c) heat capacity ratio, $C_v/T$ at pressure $p=20$ bar: (i) solid blue curve is a theoretical calculation including the pair-breaking effects of aerogel based on the theory of Ref. \onlinecite{ali11} and (ii) the purple curve is the BCS prediction for pure \He\ at the same pressure. The red shaded region at low temperature represents the contribution of the gapless excitations to the low-temperature Sommerfeld coefficient. Panel (d) shows the pressure-temperature phase diagram: (i) red curves define the bulk phase boundaries for pure \He, (ii) the data are from torsional oscillator measurements~\cite{mat97} (open circles) and acoustic impedance measurements~\cite{ger02a} (solid circles) and (iii) the solid blue curve is the theoretical prediciton for $T_c$ in aerogel with a mean-free path $\ell=140$ nm and aerogel correlation length of $\xi_a=50$ nm.
}
\end{figure}
\end{widetext}

The discovery of superfluidity in $^3$He spawned experimental and theoretical developments that have influenced many aspects of condensed matter physics.~\cite{volovik03} Concepts and phenomena discovered in studies of \He\ have been central to developments in the field of ``unconventional superconductivity'' in which one or more symmetries of the normal metallic state, e.g. rotations, reflections, time-reversal, etc., are spontaneously broken in conjunction with the $U(1)$ symmetry that is characteristic of superconductivity.\cite{mineev99}
In the case of \He\ both spin- and orbital rotation symmetries, as well as parity and time-reversal, are broken. The instability to p-wave, spin-triplet pairing allows for many possible ground states, only two of which are realized as the bulk A- and B phases of \He. Their properties are summarized in the sidebar and Fig. \ref{fig-order_parameters}.

In the context of topological quantum matter, the B-phase is the paradigm of a time-reversal invariant, three-dimensional topological superconductor, with emergent spin-orbit coupling and helical Majorana fermions confined on surfaces. Thin films of \He-A are time-reversal broken topological phases, closely related to integer quantum Hall systems, supporting chiral fermions propagating along edges of the film.\cite{volovik03,miz16,sau11}
The deep connection between the phases of \He, topological insulators and superconductors has led to wide ranging theoretical and experimental studies of \He\ in confined nano-scale geometries,\cite{lev13} notably the recent discoveries of the polar phase of \He\ nematically aligned alumina aerogel, and remarkable topological defects that possess $\nicefrac{1}{2}$ the normal quantum of fluid circulation.\cite{aut16,sau16}  

\begin{widetext}
\begin{minipage}{\textwidth}
\centering
\includegraphics[height=4cm]{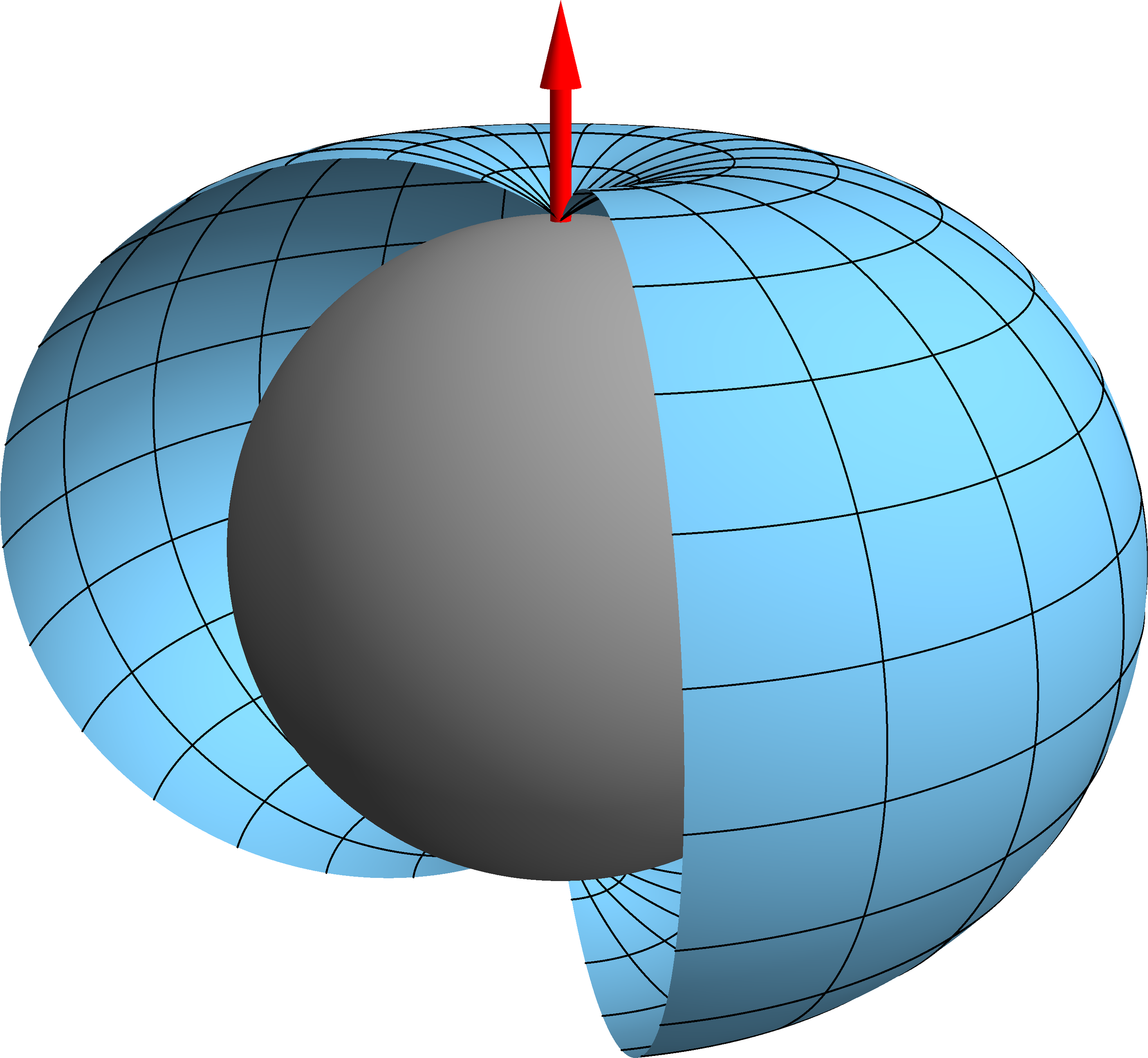}\hspace*{5mm}
\includegraphics[height=4cm]{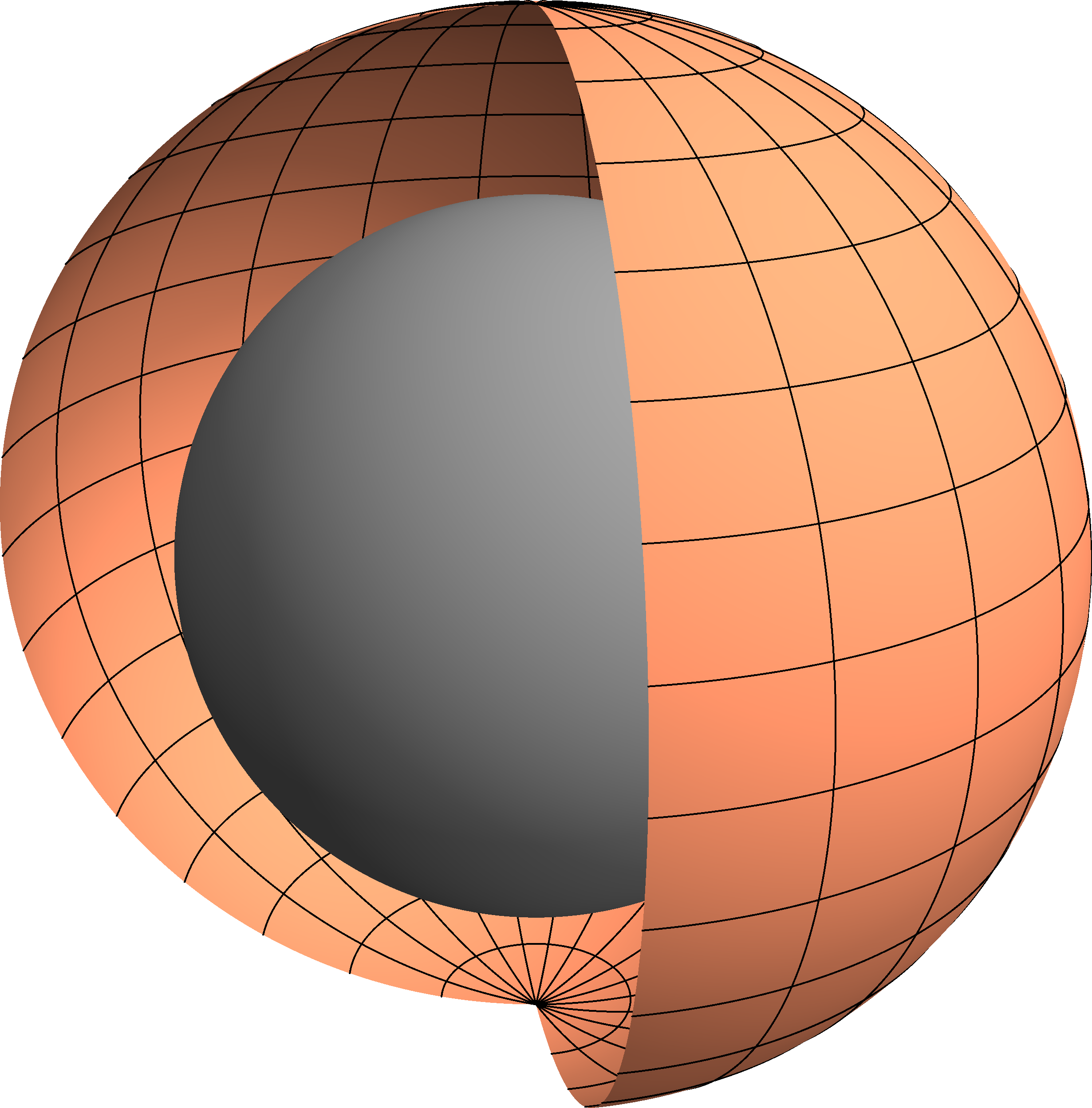}\hspace*{5mm}
\includegraphics[height=4cm]{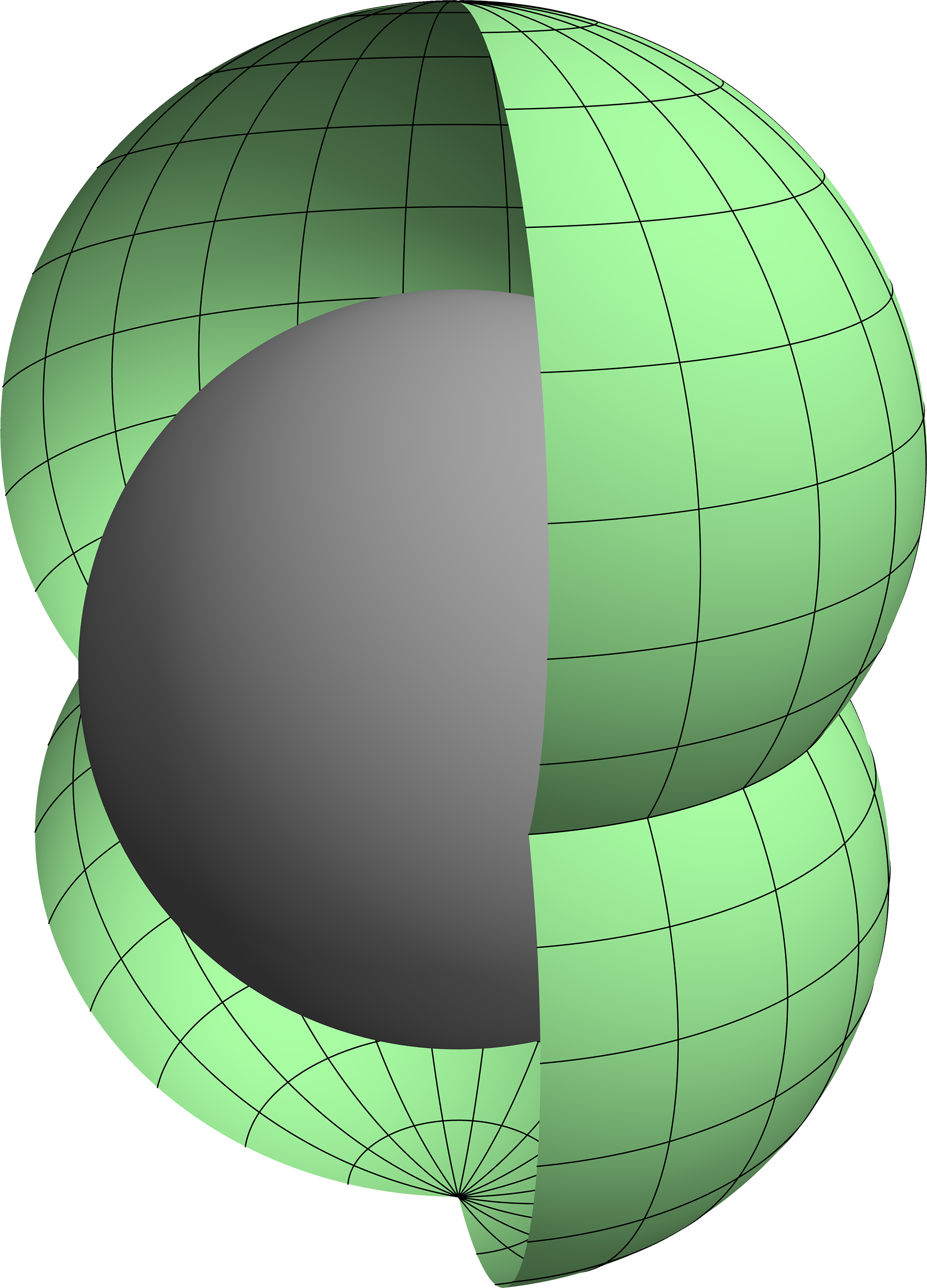}\hspace*{5mm}
\includegraphics[height=4cm]{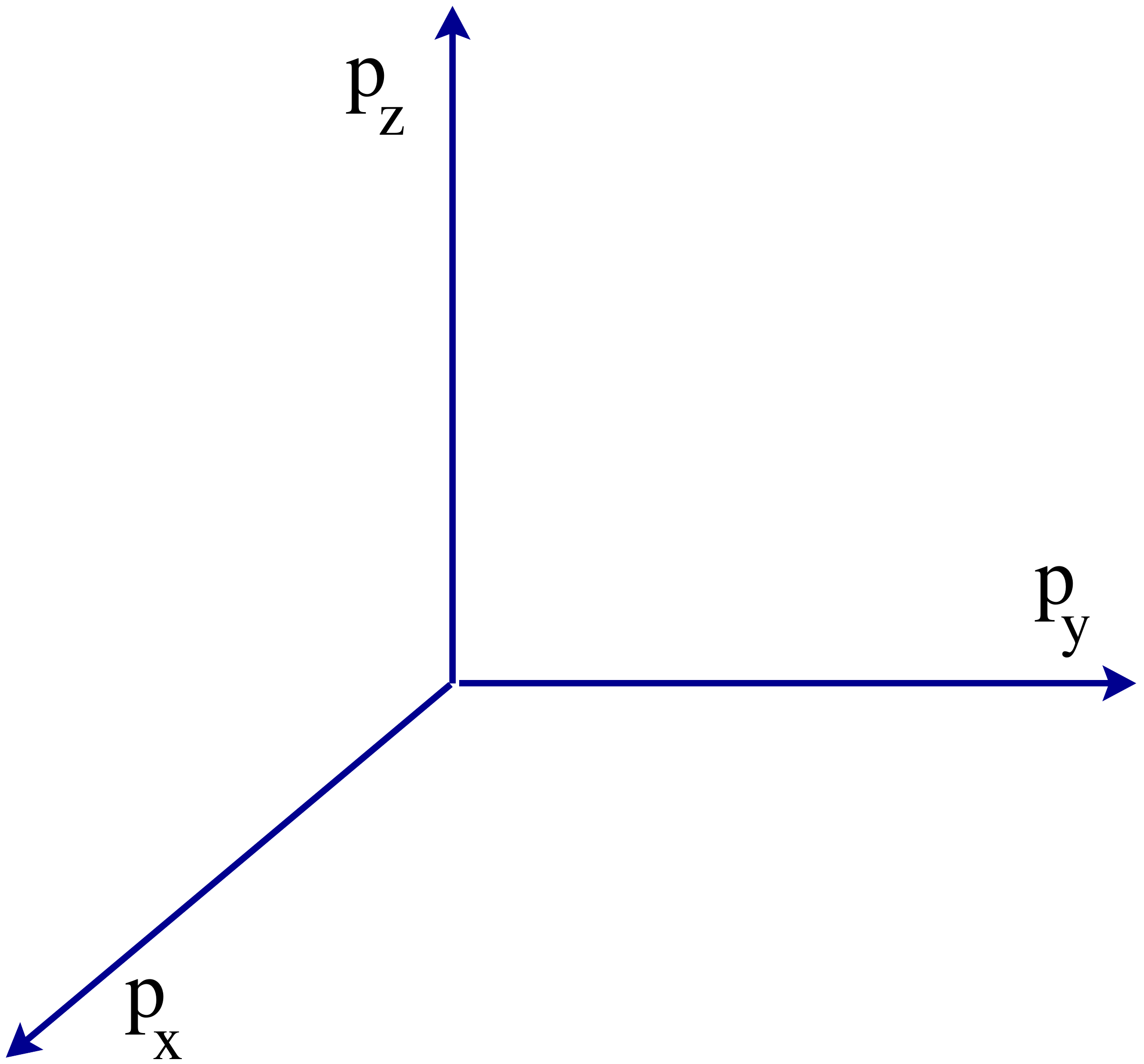}
\end{minipage}
\begin{figure}[h]
\caption{\label{fig-order_parameters} 
The Cooper pair amplitudes defined on the Fermi surface in momentum space (gray sphere) are shown from left to right for the A-phase (``chiral'' state), the B-phase (``isotropic'' state), and the newly discovered P-phase (``polar'' state) phase.~\cite{dmi15}}
\end{figure}
%
\detail{
\noindent \textbf{Phases of Superfluid $^{3}$He}\\
The superfluid phases of $^{3}$He are Bardeen-Cooper-Schrieffer condensates of p-wave ($L=1$) Cooper pairs with orbital wave functions, expressed in terms of the relative momentum, $\vp$, of constituent quasiparticles, that are linear superpositions of the p-wave states defined on the Fermi surface: $\Psi_{1,1}(\vp)=(p_x + i p_y)/\sqrt{2}$, $\Psi_{1,-1}(\vp)=(p_x - i p_y)/\sqrt{2}$, $\Psi_{1,0}(\vp)=p_z$. The Pauli exclusion principle then requires that these pairs form nuclear spin-triplet states ($S=1$). There are three superfluid phases of pure $^3$He corresponding to different realizations of the p-wave, spin-triplet manifold. Two phases, the A- and B-phases, are indicated in the phase diagram in Fig. \ref{PhaseDiagram}. A third phase, labelled A${_1}$, develops in a narrow region near T$_{c}$ in an applied magnetic field. All three phases are characterized by their nuclear spin structure and correspond to different superpositions of p-wave, spin-triplet states. The A-phase is a superposition with equal amplitudes for the oppositely polarized spin-triplet states (referred to as ``equal spin pairing'' [ESP]), $\ket{A}=\Psi_{1,1}(\vp)\left(\ket{\uparrow\uparrow}+ \ket{\downarrow\downarrow}\right)/\sqrt{2}$. The orbital wave function, $\Psi_{1,1}(\vp)$, corresponds to Cooper pairs with $L_z=+\hbar$. Thus, the A phase is ``chiral'', breaking both time-reversal and mirror symmetries. The A-phase survives in large magnetic fields without destroying Cooper pairs by conversion of $\ket{\downarrow\downarrow}$ pairs into $\ket{\uparrow\uparrow}$ pairs in order to accommodate the nuclear Zeeman energy. The A${_1}$ phase is the spin-polarized chiral state, $\ket{A_1}=\Psi_{1,1}(\vp)\ket{\uparrow\uparrow}$, and has the highest transition temperature in a magnetic field. The B-phase, which is the stable state over most of the phase diagram in zero magnetic field, is a superposition of all three triplet spin states: $\ket{B}=\Psi_{1,-1}(\vp)\ket{\uparrow\uparrow} +\Psi_{1,1}(\vp)\ket{\downarrow\downarrow} +\Psi_{1,0}(\vp)\ket{\uparrow\downarrow+\downarrow\uparrow}$. The spin- and orbital states are organized so that the B-phase is a state with total angular momentum $J=0$, i.e. the ``isotropic'' state. One of the key signatures of a B-like phase is the reduction of the nuclear magnetic susceptibility resulting from the $\ket{\uparrow\downarrow+\downarrow\uparrow}$ pairs. The B-phase is suppressed in large magnetic fields when the Zeeman energy is comparable to the binding energy of the $\ket{\uparrow\downarrow+\downarrow\uparrow}$ pairs. The transition from A to B phases is first order. The point in the phase diagram where all three phases are degenerate is the polycritical point (PCP). This singular point is destroyed by application of a magnetic field which opens up regions of stability of both the $A$ and $A_1$ phases over the full pressure range. Many of these features are fundamentally altered in the presence of disorder, notably the stabilization of the polar phase, $\ket{P} = \Psi_{1,0}(\vp)(\ket{\uparrow\uparrow}+\ket{\downarrow\downarrow})/\sqrt{2}$, which is also an ESP state like the A-phase, but with nematically aligned Cooper pairs infused into alumina aerogels with strong uniaxial anisotropy.\cite{dmi15} The momentum space amplitudes of the A- B- and P-phases are shown in Fig. \ref{fig-order_parameters}. 
}
\end{widetext}

\vspace*{-7mm}
\subsection*{Impurities and Disorder in Superfluid \He}
\vspace*{-3mm}

In almost all forms of condensed matter, impurities and structural defects are unavoidable unless special steps are taken during growth, fabrication and processing of solid state materials. Indeed for many superconducting materials, even relatively modest concentrations of chemical impurities and structural imperfections can mask their intrinsic behavior, or destroy a superconducting state with a broken rotation or reflection symmetry.

By contrast, liquid \He\ at low temperatures is the purest known form of matter. Not even \Hefour\ will dissolve in \He\ in any measureable quantity at temperatures below 10 mK. And while there are remarkable transport studies using electrons and ions forced into superfluid \He, these charged impurities are so dilute as to perturb the liquid only on nanometer scales near each ion. Thus, ion studies probe excitations of pure \He.\cite{ike13,she16} The intrinsic properties of normal and the superfluid phases is rich, well studied and quantitatively understood theoretically. For these reasons superfluid \He\ has been widely thought of as unique compared to electronic superconductors in that BCS pairing occurs in essentially a pristine material free of defects and impurity disorder. 

This changed with the discovery of superfluidity of liquid \He\ infused into silica aerogel. The ability to introduce disorder by infusing \He\ into low-density random solids, and to mechanically or chemically modify the properties of the random medium, opened wide ranging opportunities for studying the effects of disorder on an unconventional BCS condensate for which the intrinsic properties are quantitatively understood.  

\begin{widetext}
\begin{minipage}{\textwidth}
\hspace*{-5mm}\begin{minipage}{0.33\textwidth}
\vspace*{-60mm}
\includegraphics[width=\textwidth]{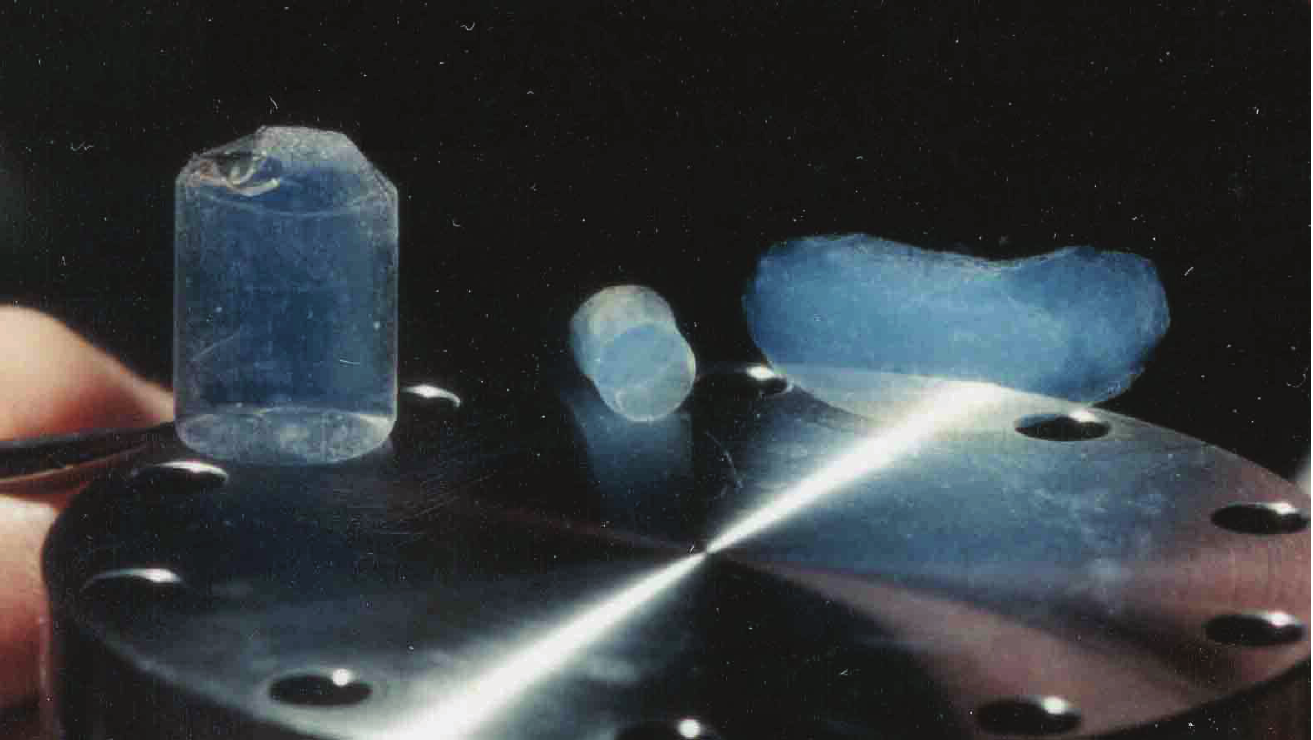}
\includegraphics[width=\textwidth]{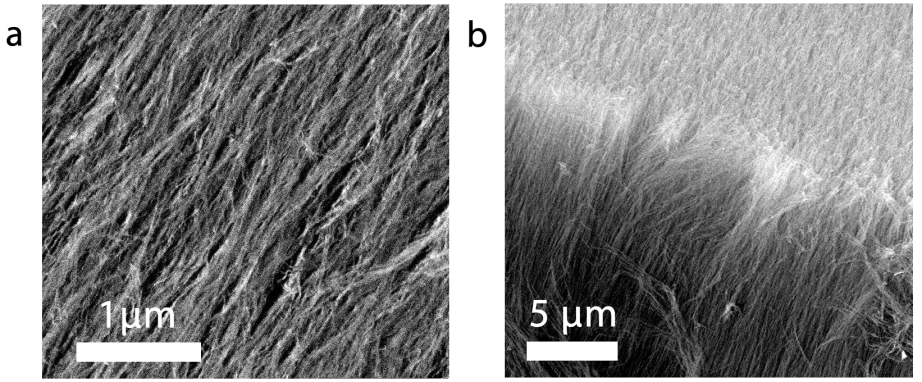}
\end{minipage}\hspace*{2mm}
\includegraphics[width=0.33\textwidth,angle=00]{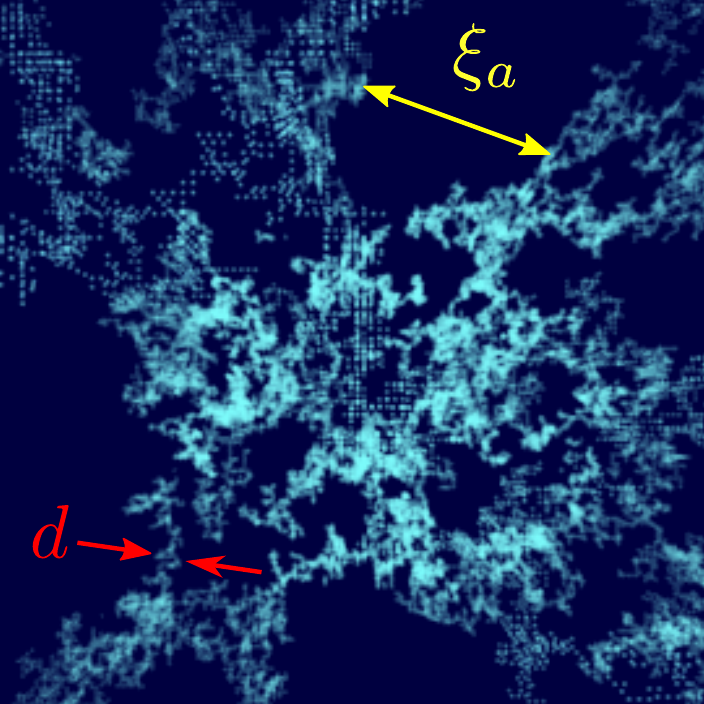} \hspace*{1mm}
\includegraphics[width=0.32\textwidth,angle=00]{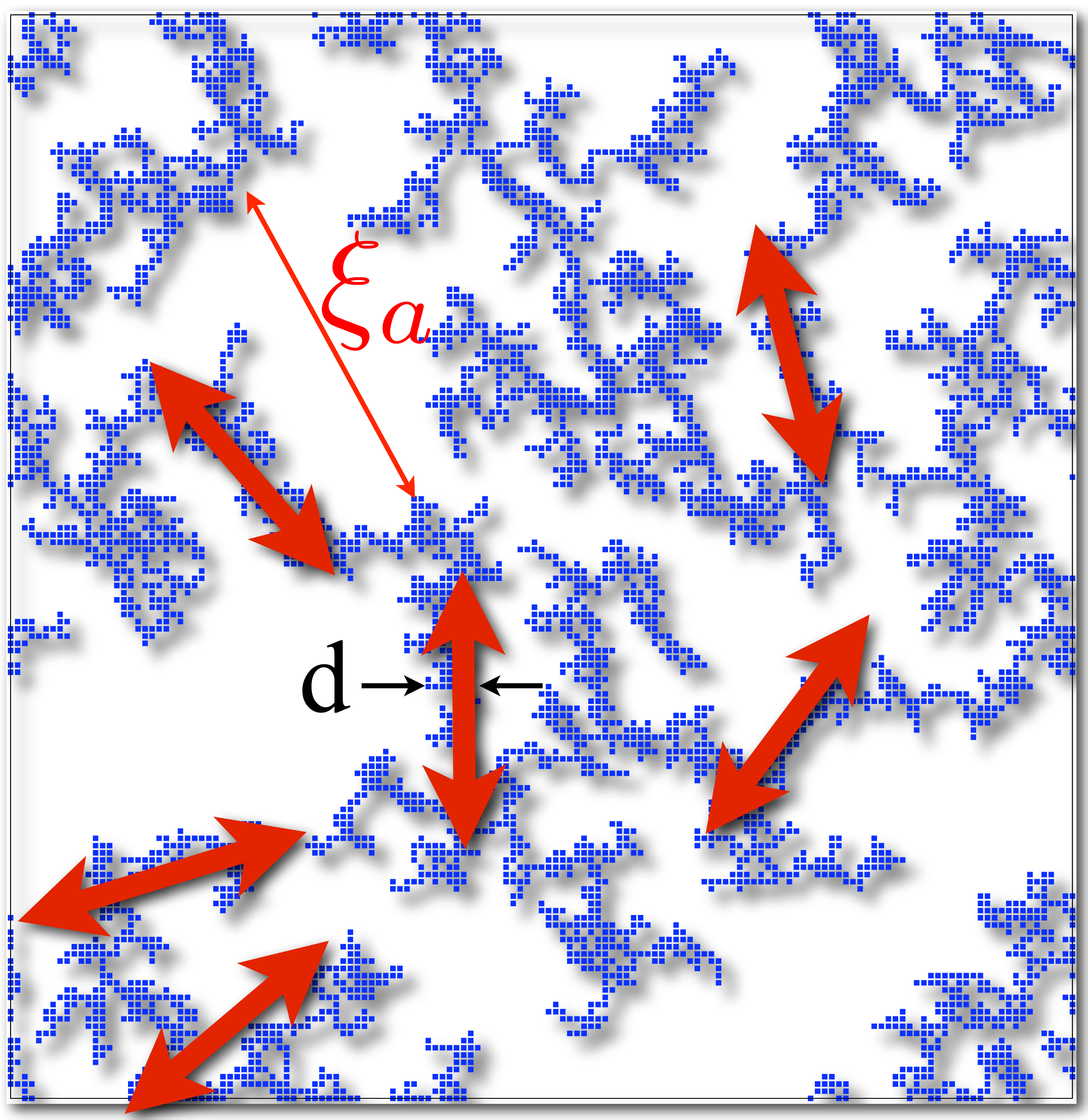}
\end{minipage}
\begin{figure}[h]
\caption{\label{Aerogel}
Left panel, top: Photograph of three silica aerogel samples with porosities 95\,\%, 98\,\%, and 99\,\% from left to right, placed on the top of a 13\,cm diameter high pressure autoclave used to supercritically dry the samples. The light blue color is due to Rayleigh scattering. Left panel, lower: A nematic aerogel of alumina stands can stabilize the polar phase of nematically aligned Cooper pairs. Middle panel: A 3D DLCA simulation of 98\% silica aerogel showing the fractal structure of the \sio\ strands. Right panel: A 2D DLCA simulation illustrating the strand size, $d$, aerogel correlation length, $\xi_a$, and the local anisotropy of the aerogel.
}
\end{figure}
\end{widetext}

\vspace*{-5mm}
\subsection*{Silica Aerogel}
\vspace*{-3mm}

Silica aerogels are low-density solids with porosities up to 99.5\,\% by volume. The structure is that of a dilute network of thin \sio\ strands and clusters with typical dimensions of 2 to 5\,nm. Aerogels are fascinating materials that have found applications ranging from Cerenkov counters in particle physics to light-weight, transparent thermal insulation for the space vehicles.~\cite{fri92} 
The top left panel of Fig.~\ref{Aerogel} shows a photograph of three silica aerogels with porosities 95\%, 98\% and 99\% from left to right. The aerogels are transparent; consequently, we infer that the structure is homogeneous on length scales of order the wavelength of visible light. The blue color is caused by Rayleigh scattering of light by the density fluctuations of the aerogel.

Silica aerogel is formed from a base-catalyzed synthesis of silica clusters approximately 3\,nm in diameter. Gelation is performed from tetramethylorthosilicate dissolved in alcohol. Small clusters of \sio\ aggregate to generate strands and clusters that bond to form the gel structure. The wet gel is dried at a supercritical pressure in a high pressure autoclave to avoid collapse of the microstructure from capillary forces at the liquid-gas interface. The resulting material is air stable and hydrophobic. 
Samples grown using a one-step gelation procedure~\cite{por99} were found to be uniformly isotropic with no identifiable optical axis. Small-angle X-ray scattering measurements~\cite{por99} show fractal correlations of the aerogel density, which is characteristic of the process of diffusion-limited cluster aggregation (DLCA).~\cite{mea83} The power-low correlations span two or more decades in wavevector for the higher porosity aerogels, terminating for $q \approx q_a=\pi/\xi_a$, $\xi_a\approx 30-50$\,nm is a correlation length that is a measure of the largest open voids within the aerogel. On longer length scales statistical self-similarity halts, and the aerogel becomes isotropic and homogeneous. 
These results are supported by numerical DLCA simulations of a 98\,\% porosity aerogel. The center panel of Fig.~\ref{Aerogel} shows the image of the 3D structure at a length scale of several aerogel correlation lengths. The largest open voids correspond to a radial distance of order $\xi_a$. On shorter length scales, self-similar open regions characteristic of a random fractal are clearly visible. The right panel of Fig.~\ref{Aerogel} shows a 2D simulation highlighting the short-range, local anisotropy and characteristic dimensions of the \sio\ strands, $d\approx 2\,\mbox{nm}$.

While it was known in the early 1990s that the Bose superfluidity of \Hefour, with an atomic-scale correlation length, could survive the random potential of porous glass and silica aerogel, the discovery of superfluidity of \He\ infused into silica aerogel was a surprise. The superfluid phases of pure \He\ were known to be fragile, onsetting three order of magnitude lower in temperature. Scattering by the aerogel structure could easily destroy the weakly bound, p-wave Cooper pairs with radial extent of order $\xi\approx 50\,\mbox{nm}$. Given the characteristic length scale of $\xi_a\approx 30\,\mbox{nm}$ in higher porosity silica aerogels it was unclear whether or not the orbital correlations that give rise to superfluidity in $^3$He could survive within the random network of strands and clusters that make up silica aerogel.

What can also be inferred from the open structure of the DLCA simulated aergel is that there is an even longer length scale corresponding to the geometric mean free path (mfp) for point particles moving along straight trajectories within the open regions of the aerogel. For 98\% porous aerogels the mfp is determined from the DLCA simulation to be $\ell\approx 180$ nm, much longer than the aerogel correlation length, reflecting a fractal structure with a probability distribution of free paths characteristic of a L\'evy distribution.~\cite{fog98}
One expects that the transport mean-free-path for \He\ quasiparticles moving at the Fermi velocity will be the same. Indeed, this is borne out by analysis of measurements in normal \He\ infused into 97-99\,\% porous aerogels, including spin diffusion, sound attenuation, and heat transport,~\cite{sau05,sha01,rai98,sau10} from which transport mean-free-paths are measured to be $\ell\approx 100-200\,\mbox{nm}$ depending on the porosity. The long mfp and the fractal structure of the aerogel are key to a theoretical understanding of superfluidity of \He\ confined in random solids. The agreement between the mfp obtained from DLCA simulations and measurements provided impetus for the application of scattering theory to describe the effects of aerogel on the superfluid phases of $^3$He.\cite{thu98}

\vspace*{-5mm}
\subsection*{Superfluid Fraction}
\vspace*{-3mm}

Superfluid \He\ infused into silica aerogel was discovered with a torsional oscillator - a torsion rod attached perpendicular to a disk containing the \He-aerogel sample. The resolution of high-Q torsional oscillators is ideal for measurements of superfluid flow.~\cite{por95} In the two-fluid model, the superfluid state of \He\ is the superposition of a normal fluid of excitations, viscously clamped to the porous structure, and an inviscid superfluid of Cooper pairs. The latter decouples from the aerogel and does not contribute to the moment of inertia of the mechanical oscillator. At the onset of superfluidity the resonant period of the oscillator decreases. The decrease in moment of inertia is directly related to the superfluid fraction, shown in Fig.~\ref{PhaseDiagram}a). However, in contrast with pure superfluid $^3$He, the superfluid fraction as $T\rightarrow 0$ is much less than unity, reflecting the disorder-induced spectrum of gapless excitations.

\vspace*{-5mm}
\subsection*{NMR Frequency Shift}
\vspace*{-3mm}

In parallel with the measurements of superfluid fraction, NMR measurements detected the superfluid phase of \He\ infused into silica aerogel from the NMR frequency shift, shown in Fig.~\ref{PhaseDiagram}\,(b).~\cite{spr95} The NMR absorption spectrum for the superfluid phases ($\alpha =$ A, B, P) is centered at a frequency $\omega$ that is shifted away from the Larmor frequency, $\omega_{\text{L}} = \gamma B$, by an amount $\Delta\omega$,~\cite{leg72}
\begin{equation}
\Delta\omega \equiv \omega-\omega_{\text{L}} \simeq F_{\alpha}(\beta)
                    \frac{\Omega^{2}_{\alpha}}{2\omega_{\text{L}}}
\,,
\end{equation}
for magnetic fields well-above the nuclear dipolar field, $\Omega_{\text{A}}/\gamma \approx 3\,\mbox{mT}$. The transverse shift is determined by the longitudinal resonance frequency, $\Omega_{\alpha}$, and a function $F_{\alpha}(\beta)$ that depends on the relative spin-orbit orientation of the Cooper pairs, and the magnitude of the r.f. excitaiton expressed in terms of the tipping angle, $\beta$. The function $F_{\alpha}$ provides a ``fingerprint'' of each superfluid phase: for small tipping angles, $F_{\text{B}}\approx 0$, $F_{\text{A}}\approx 1$, $F_{\text{P}}\approx 4/3$. In the Ginzburg-Landau limit, $T\lesssim T_c$, the longitudinal resonance frequencies are proportional to the amplitude of the order parameter for each phase, and thus the transverse shift is linear in temperature near $T_c$.~\cite{leg72} Fig.~\ref{PhaseDiagram}\,(b) shows the onset of a frequency shift, which is similar to that of the superfluid $A$-phase in pure \He, but much smaller in magnitude. The reduction of the magnitude of the NMR frequency shift compared to pure $^3$He, similar to that for the superfluid fraction, is a reflection of the pair-breaking effect of scattering by the aerogel strands.~\cite{thu98}

\begin{figure}[h]
\includegraphics[width=80mm]{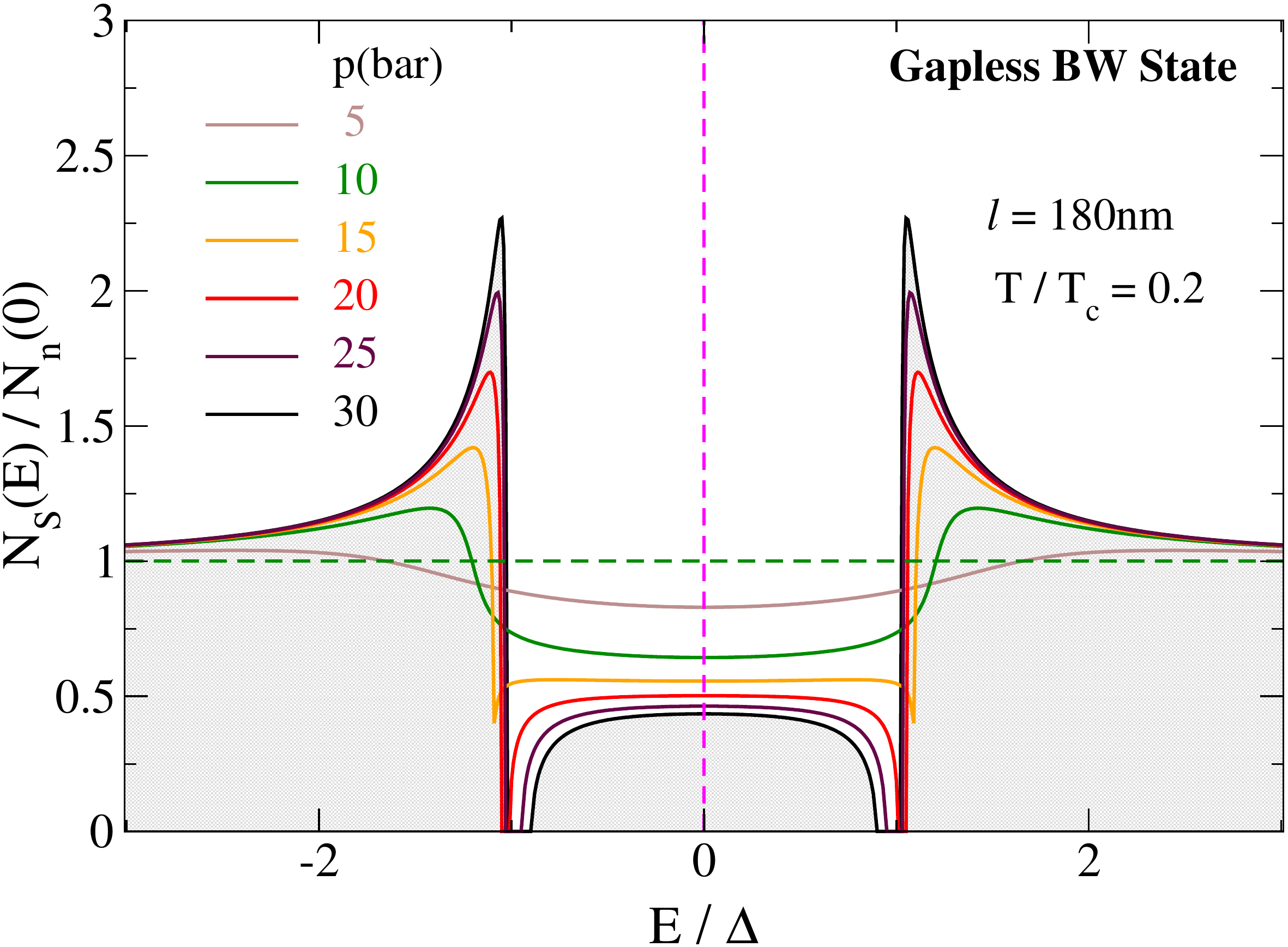}
\caption{\label{DOS}
The density of states normalized to that of normal \He\ at the Fermi level for the B phase in aerogel with porosity $\approx 98.5\%$, mean-free-path $\ell=180$ nm, at pressures ranging from $p=5-30$ bar. Note that $N_{\mbox{\tiny S}}(0)\ne 0$, i.e. the B-phase in aerogel is gapless at any pressure.~\cite{sha01}
}
\end{figure}

\vspace*{-5mm}
\subsection*{Gapless Superfluidity}
\vspace*{-3mm}

Pure \Heb\ is a fully gapped BCS superfluid. At low temperatures, $T\ll\Delta/k_{\text{B}}\approx 2\,\mbox{mK}$, where $\Delta$ is gap in the excitation spectrum, there is essentially zero normal fluid. The heat capacity, thermal conductivity and normal fluid fraction are exponentially suppressed by the gap in the quasiparticle spectrum. NMR measurements show that the B-phase in isotropic silica aerogels has the same spin-orbit symmetry as that in pure \Heb, however, the low-temperature thermodynamic and transport properties of the superfluid B phase in isotropic silica aerogels are radically different than their counterparts in pure \Heb.

Thermal conductivity ($\kappa$) and heat capacity ($C_{\text{V}}$) measurements,~\cite{fis03,cho04} show that the low-temperature limits for $\kappa$ and $C_{\text{V}}$ of superfluid \He\ in aerogel are linear in temperature. The experimental data shown in Fig. \ref{PhaseDiagram}c, combined with the third law of thermodynamics, \emph{requires} that $\lim_{T\rightarrow 0} C_{\text{V}}/T \propto N_{\text{S}}(0) \ne 0$. A similar conclusion is obtained from analysis of measurements of the thermal conductivity. The results are consistent with the theoretical prediction that superfluid \He\ infused into high porosity aerogel is a ``gapless superfluid'' with a finite density of states at the Fermi energy over the entire pressure range, i.e. that a normal-fluid component coexists with the superfluid condensate in the $T\rightarrow 0$ limit, as shown in Fig.~\ref{DOS} for the density of states,\cite{sha01} and in Fig. \ref{PhaseDiagram}(c) for the Sommerfeld constant $\lim_{T\rightarrow 0}C_{\text{V}}/T$ obtained from theoretical analysis of the heat capacity of superfluid \He\ infused in 98\% silica aerogel.~\cite{cho04,ali11}

\begin{widetext}
\vspace*{-4mm}
\includegraphics[width=0.45\columnwidth]{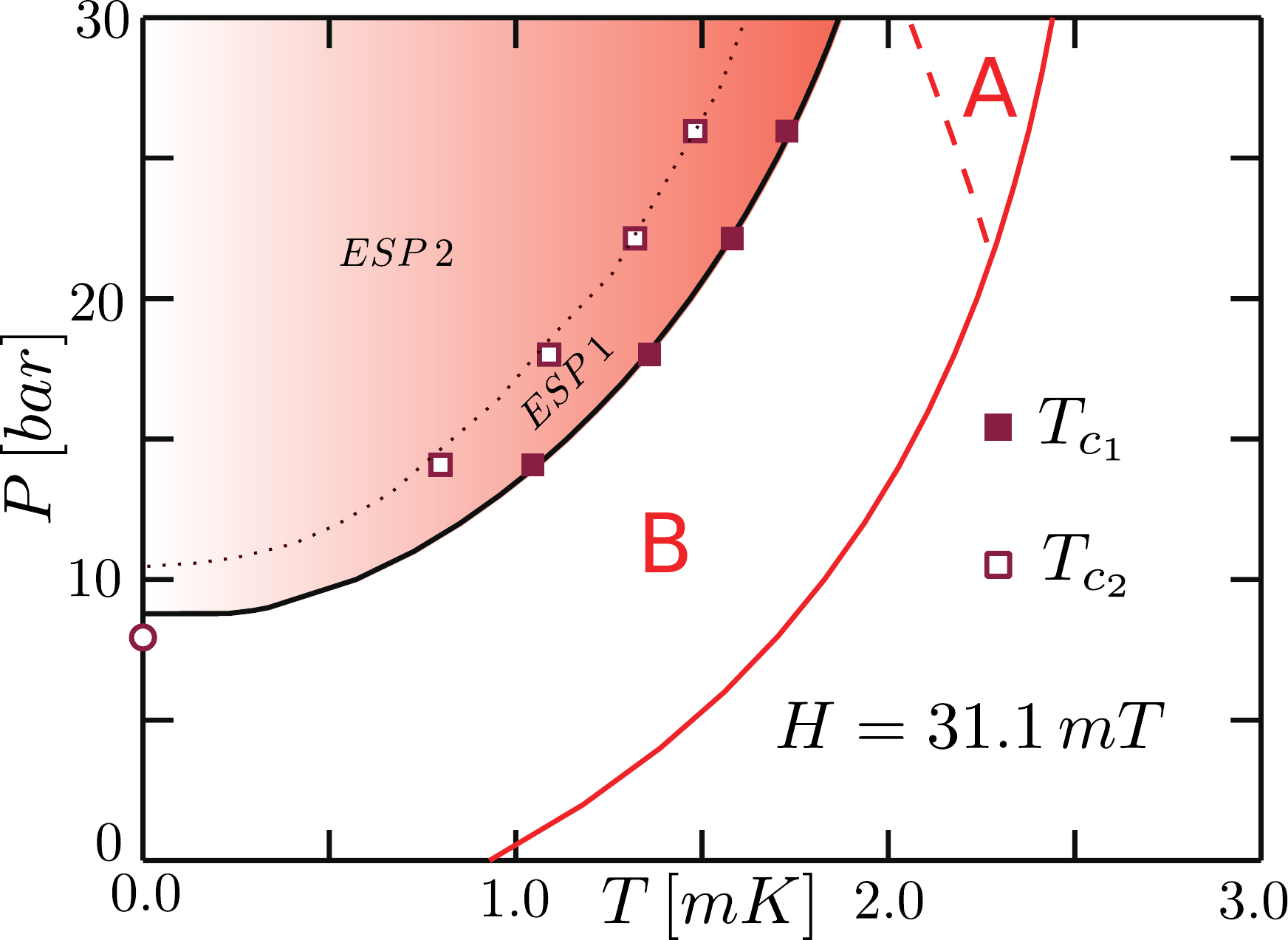}
\includegraphics[width=0.50\columnwidth]{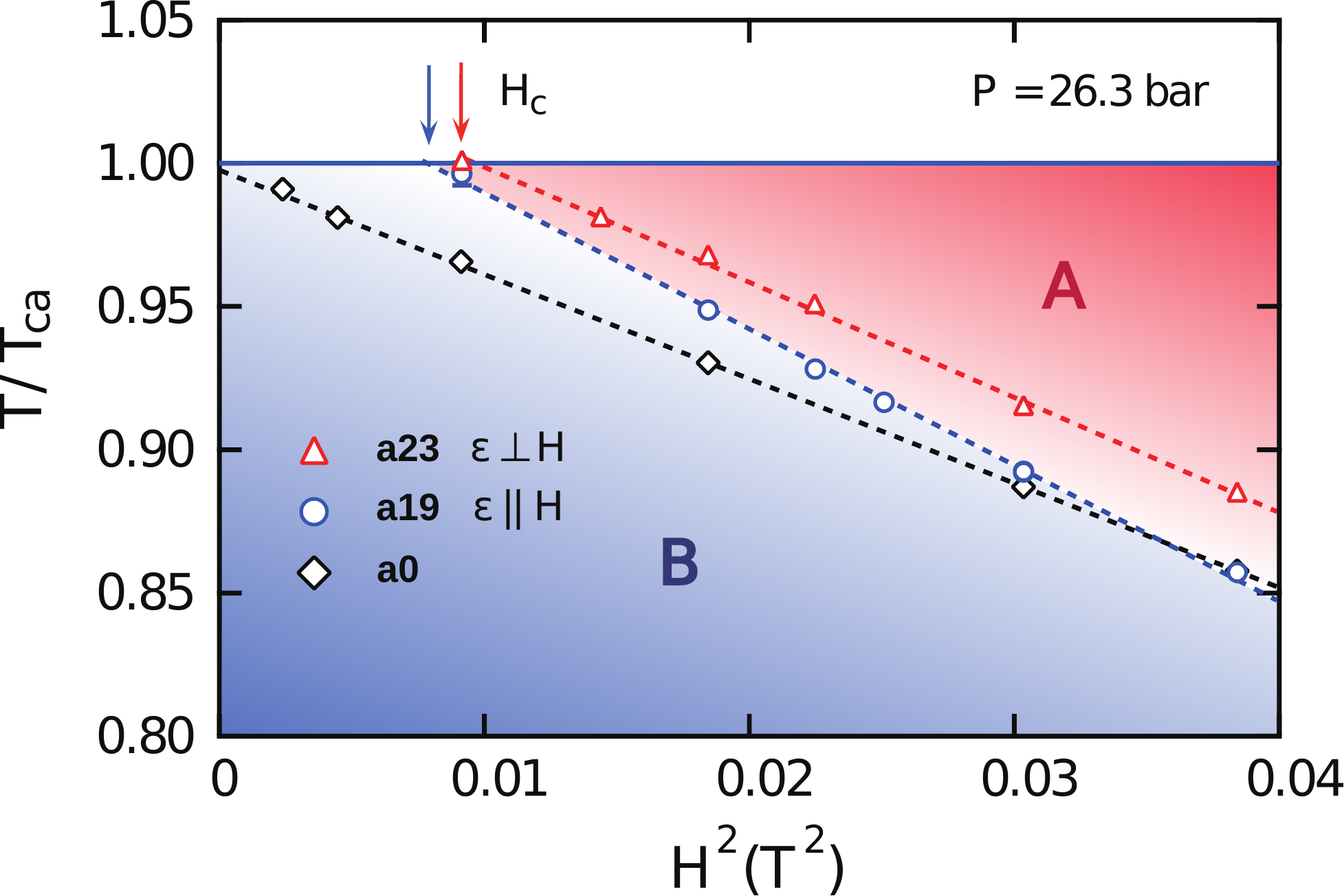}
\begin{figure}[h]
\caption{\label{AnisotropicPDs} Phase diagrams for superfluid \He\ in anisotropic silica aerogels. Left panel: p-T phase diagram for stretched aerogel with strain $\varepsilon = 0.14$. Two ESP states are observed by NMR. ESP-1 is the A-phase with a chiral axis along the strain. A transition to a second ESP state with lower symmetry occurs at $T_{c_2}$. Right panel: T-H phase diagram for compressed aerogel with strain $\varepsilon \approx -0.20$. Compression stabilizes the B phase at a critical field, $H_c\approx 0.1\,\mbox{T}$.}
\end{figure}
\end{widetext}

\vspace*{-5mm}
\subsection*{Isotropic Silica Aerogel - The Gapless B Phase}
\vspace*{-3mm}

Based on NMR and acoustic attenuation measurements the equilibrium superfluid phase in zero applied field, in isotropic aerogels, is the gapless $B$-phase at all pressures.~\cite{ger02,pol11,pol12} 
The A phase is stabilized in small magnetic fields as shown in the right panel of Fig. \ref{AnisotropicPDs} for isotropic aerogel (labelled ``a0''). The A-B transition is first order and depends on the square of the magnetic field as a result of the depairing effect of the Zeeman energy for the B phase. The zero-field stability of the B phase in isotropic aerogel is evident in Fig.~\ref{AnisotropicPDs} from the extrapolation of the equilibrium A-B transition line, measured on warming, to zero magnetic field. These results confirmed the prediction that scattering within in a isotropic aerogel favors the isotropic B phase.~\cite{thu98}

However, on cooling in zero applied field, a \emph{metastable} $A$-phase is observed, which is absent on warming from the B phase. The implication is that rarefied regions (static fluctuations in density) within the aerogel are favorable at high pressures to nuclei of A phase which then supercool as a metastable A phase. Once the metastable A phase collapses to B, the A phase is no longer accessible on warming.

\vspace*{-5mm}
\subsection*{Anisotropic Aerogels - New Phases}
\vspace*{-3mm}

Anisotropic silica aerogels can be grown or engineered using several techniques. Stretching or compressing silica aerogel introduces long-range uniaxial anisotropy. In contrast to the B phase in isotropic aerogels the superfluid phases and the phase diagrams in anisotropic aerogels are radically different.~\cite{pol11,pol12} Stretching is achieved by a growth process with an elevated amount of catalyst, resulting in uniform radial shrinkage during supercritical drying.~\cite{pol12} Stretched anisotropy with strain of order 14\,\%, stabilizes an A-like phase labelled ESP-1 in the left panel of Fig. \ref{AnisotropicPDs}.~\cite{pol12,sau13} This phase is stabilized by anisotropic scattering.~\cite{thu98,sau13} NMR measurements~\cite{pol12,li14} with magnetic fields oriented both parallel and perpendicular to the strain axis identify ESP-1 as a chiral A-like phase with the angular momentum aligned along the strain axis, consistent with theory.~\cite{sau13} A second transition, also an ESP state (ESP-2), onsets at lower temperature, $T_{c_2}$ (open squares), which is theoretically predicted to be a polar-distorted, bi-axial chiral phase.\cite{sau13} Indeed measurements in the lower temperature phase demonstrate that the angular momentum orients off the strain axis.~\cite{li14}

Anisotropy is also created by compressing isotropic aerogels {\it in situ} within a glass tube after drying. Varying degrees of strain have been achieved in samples of 98\,\% porous silica aerogel. The uniformity of the strain that is induced is quantified using optical birefringence.~\cite{li15} Anisotropy introduced by compression leads to dramatic effects on the superfluid state and the phase diagram.

An unexpected feature is that the non-ESP B-phase is more stable than the ESP A phase in low magnetic fields, as shown in the right panel of Fig.~\ref{AnisotropicPDs}. A new critical field, $H_c$, appears only for \He\ in compressed aerogels, corresponding to a tri-critical point between the normal, A- and B-phases. The critical field is proportional to the strain in the aerogel, with $H_c = 0.1\,\mbox{T}$ for 20\,\% strain at $p = 26$\,bar.~\cite{li15} 

The region of very low magnetic field has been explored by torsional oscillator techniques, where a new phase is found very close to $T_c$. Measurements of the superfluid fraction of this new phase are consistent with a polar phase of Cooper pairs nematically aligned with the strain axis of the compressed aerogel.~\cite{ben11,sau13}

\vspace*{-5mm}
\subsection*{Nematic Alumina Aerogel - The Polar Phase}
\vspace*{-3mm}

A new class of aerogels has been produced from the growth of oriented Al$_2$O$_3$ strands (see left, lower panels of Fig. \ref{Aerogel}). ``Nematic aerogels'' called \emph{nafen} can be grown with extreme anisotropy. The degree of anisotropy has been determined from spin-diffusion measurements; ratios of $8:1$ are obtained for the ballistic mfp of quasiparticles moving along or transverse to the nematically aligned strands.\cite{dmi15b} Superfluidity of \He\ in nematic aerogels is found in both NMR and torsional oscillator studies.\cite{dmi15,zhe16} In particular there is evidence from both measurements that the first transition from the normal phase is to the polar state in which the Cooper pairs condense into a single p-wave orbital aligned along the nematic axis of nafen. The observation of a polar state is significant for several reasons. The polar state does not occur in pure superfluid \He, but was predicted theoretically to be stabilized in strongly anisotropic aerogels and channels.\cite{aoy06}\cite{sau13,wim15} Secondly, the polar state was recognized to be a favorable candidate to support the elusive half-quantum vortex (HQV) that was first predicted theoretically for p-wave, spin-triplet superfluids more than 40 years ago.\cite{vol76} Not long after the discovery of the polar phase came the announcement by the Helsinki-Moscow collaboration of the detection of HQVs by NMR spectroscopy on the superfluid polar phase of \He\ confined in nafen aerogel in a rotating sample cooled to $T=300\,\mu\mbox{K}$.~\cite{aut16}

\vspace*{-5mm}
\subsection*{Topological Superfluid $^3$He}
\vspace*{-3mm}

The discoveries of new phases of \He\ confined in high-porosity disordered solids provide insights into the effects of disorder on superfluids and superconductors with complex symmetry breaking, and has opened new research directions into topological phases confined in nano-scale materials. Studies range from the physics of topological excitations - Majorana and Weyl Fermions confined on boundaries, interfaces and topological defects,~\cite{chu09,vol09a,miz12,oku12,sau11,tsu12,wu13,ike15,tsu17,ike18} to the nucleation of topological defects in broken symmetry phases during non-equilibrium phase transitions.~\cite{elt10,aut16}

The knowledge obtained from research on superfluid \He\ confined in aerogels, combined with theoretical developments in topological quantum matter, has led to new directions in the study of quantum fluids confined in engineered nano-scale structures,~\cite{lev13,lev14} from high-Q fluid-mechanical resonantors, to nano-fluidic devices~\cite{zhe17} and ultrasonic cavities for quantum transport and dynamics of quantum fluids.~\cite{bra16,zhe16a,zhe17a,byu18} This marriage of quantum fluids, nano-scale materials and nano-fabrication is part of a new frontier in the field of hybrid quantum materials.~\cite{lee17}

\subsection*{Acknowledgements}
\vspace*{-3mm}

The research of the authors is supported by the National Science Foundation: Grants DMR-1106315 and DMR-1508730 (Sauls), DMR-1602542 (Halperin) and DMR-1708341 (Parpia).
A draft of this article was written at the Aspen Center for Physics, which is supported by National Science Foundation Grant PHY-1607611.


\end{document}